# Towards 6G-V2X: Aggregated RF-VLC for Ultra-Reliable and Low-Latency Autonomous Driving

Gurinder Singh, Anand Srivastava, Vivek Ashok Bohara, Md Noor-A-Rahim, Zilong Liu, and Dirk Pesch

***Abstract*—We are witnessing a transition to a new era of mobility where pervasively connected (semi-)autonomous cars will deliver significantly improved safety, traffic efficiency, and travel experiences. A diverse set of advanced vehicular use cases such as platooning, remote driving, and fully autonomous driving will be made possible by building on emerging sixth-generation (6G) wireless networks. Among many disruptive 6G wireless technologies, the principal objective of this paper is to introduce the potential benefits of the hybrid integration of Visible Light Communication (VLC) and Radio Frequency (RF) based Vehicle-to-Everything (V2X) communication systems. We examine the impact of interference as well as various meteorological phenomena viz. rain, fog and dry snow, on the proposed Link aggregated (LA) aided hybrid RF-VLC V2X systems. The simulation results suggest that our proposed LA-aided hybrid RF-VLC V2X systems have the potential to achieve a high level of reliability (estimated at approximately 99.999%) and low latency (potentially less than 1 ms) within a range of up to 200 m, even in scenarios affected by interference and adverse meteorological conditions. To stimulate future research in the hybrid RF- VLC V2X area, we also highlight the potential challenges and research directions.**

## I. Introduction

CONNECTED autonomous vehicle (CAV) technologies are expected to support improved road safety, traffic efficiency, and driving comfort for a future society in motion. To fully support CAV, there is a growing demand for ultra-reliable and low-latency exchange of sensing and control data collected by the many onboard sensors and communication devices in a modern car. Such a demand is believed to be met by the next-generation vehicle-to-everything (V2X) communication technologies in the form of, for example, vehicle-to-vehicle (V2V) and vehicle-to-infrastructure (V2I) communications [1].

There are two major V2X streams: dedicated short-range communication (DSRC) based V2X and cellular-V2X (C-V2X). While DSRC represents a mature, cost-efficient V2X technology, C-V2X has attracted much attention in recent years because of its significantly improved coverage, throughput, and latency. The latter is due to the improved physical layer, centralized resource allocation, as well as sophisticated cellular infrastructure. In the 3rd generation partnership project (3GPP), several V2X initiatives, such as LTE-V2X and 5G New Radio (NR)-V2X, have contributed to the prominence of C-V2X.

Gurinder Singh is with the Department of Communication and Computer Engineering, The LNM Institute of Information Technology, Jaipur, Rajasthan, India-302031 (e-mail: gurinder.singh@lnmiit.ac.in).

Anand Srivastava and Vivek Ashok Bohara are with Centre of Excellence on LiFi, IIIT-Delhi-110020, India (e-mail:anand@iiitd.ac.in and vivek.b@iiitd.ac.in)

Md Noor-A-Rahim and Dirk Pesch are with the School of Computer Science & IT, University College Cork, Cork, T12 K8AF Ireland (e-mail: m.rahim@cs.ucc.ie, d.pesch@cs.ucc.ie).

Zilong Liu is with the School of Computer Science and Electronics Engineering, University of Essex, Colchester CO4 3SQ, U.K. (e-mail:zilong.liu@essex.ac.uk).

Despite an abundance of both advanced sensors and communication devices, however, new challenges arise for the next-generation V2X networks. More explicitly, stringent requirements on system reliability ($\geq$ 99.999%), end-to-end latency (<1 ms), coverage-quality, spectral efficiency, energy rating, networking, and privacy/security specifications should be met to support a wide range of C-V2X use cases [2]. Although the current C-V2X technology (such as 5G-NR-V2X) offers substantial performance gains over its predecessor, the improved performance is achieved at the cost of requiring additional spectrum and hardware resources while utilizing LTE-based system architectures and mechanisms. In view of this, it is believed that 6G-V2X will be a prominent supporter of the evolution towards a truly Intelligent Transportation System (ITS) and the realization of the emerging CAV technology by addressing the limitations of the current C-V2X systems.

Departing from conventional communication networks, a paradigm shift in favour of more efficient and highly flexible V2X communications is necessary. In fact, this transformation is beginning to take shape with the intensifying research into 6G that incorporates a number of disruptive concepts [1]. In addition to intelligent and ubiquitous V2X connectivity, 6G is expected to provide significant data rate increases (e.g., up to Tbps), extremely fast wireless access (e.g., in the sub-milliseconds range), a massive increase of connection density (e.g., $10^7$ devices/km$^2$ or higher), as well as more extensive, more energy-efficient, and more environmentally friendly three-dimensional (3D) communications through the emerging aerospace integrated networks (e.g., with the aid of satellites, high-altitude platforms, unmanned aerial vehicles). To realize the above grand vision of 6G-V2X, this paper advocates the intrinsic amalgamation of Radio-Frequency (RF) and Visible Light Communication (VLC) technologies which are *complementary* to each other due to their respective strengths.

***Background:*** The integration (and co-existence) of VLC-based V2X communications with the classic RF-based communications is useful for increasing the data rates, reducing the transmission latency, improving the reliability, reducing the power consumption, and enhancing the driving safety. Such integrated systems can be used to enable many V2X use cases. Targeting at supporting future CAVs, for instance, one requires close monitoring of the surrounding areas around the vehicle by deploying various sensors such as LIDAR, RADAR, camera, ultrasonic sensor, etc. In this case, a camera can not only allow the vehicle to recognize its surrounding objects but it can also be used as a transceiver for VLC. Such a camera can then communicate via VLC with a large number of devices because its spatial separation feature can facilitate, for example, object recognition around the vehicle and obstacle location estimation. Moreover, such an integrated system is very useful at road intersection, where detecting vulnerable road users in blind-spots is a challenging task especially when the line-of-sight (LoS) path is obstructed [3], [4].

While the current 5G standard does not include VLC as an integrated technology, 6G-V2X is expected to include and integrate VLC as an enabling technology for vehicular networking applications. VLC is considered as a technology that supports 6G because of its very large bandwidth and operation at THz frequencies [5]. The extra-wide bandwidth is expected to achieve higher data rates than 5G technology. Since VLC can utilize the existing LED lights that are already being used for illumination, it is more energy efficient and economically sustainable solution. Further, VLC transceiver design is less complex than RF based systems, because of much less severe multipath effect. In the literature, the authors in [6] have also shown that a hybrid RF-VLC system an achieve device cost saving as high as 47% and reduced power consumption by 49%, when it is compared to LED lights with RF technology. A hybrid RF-VLC solution could be used to reduce the high density of RF users by switching to VLC connections where possible and, in reverse, to increase the range limitation of VLC technology. However, one of the most critical issues for vehicular-VLC (V-VLC) arises from its outdoor operation. In particular, meteorological phenomenona such as fog, rain, snow, etc., can significantly influence the reliability and range of V-VLC as reported in [7]. Therefore, it is necessary to study and analyze the impact of those meteorological phenomena on the performance of hybrid RF-VLC V2X communications, qualitatively and quantitatively.

***Practical Applications:*** From an industry standpoint, exciting advances have been made in implementing and commercializing VLC technology to create new value chains. Several pilot projects have been launched to promote research in this new field. For instance, pureLiFi is the global leader in bringing to market the world's first commercial light antennas which makes light fidelity (LiFi) possible for a variety of applications from smart cars to smartphones. Besides, various globally recognized LiFi companies such as Oldecomm, VLNComm, Panasonic, Signify, and Lucibel also focus on bringing next-generation VLC front-end products to commercial markets[1]. In particular, Oledcomm, a French telecommunications company, focuses its research into LiFi systems that can be used in the automotive sector.

In this article, we review how appropriate link aggregation of V-VLC and V-RF improves the network performance as compared to standalone RF or VLC based V2X communication systems under different meteorological factors at road intersections. Link aggregation results in more efficient use of physical resources and improved reliability and availability. To the best of our knowledge, this is the first work that presents an in-depth discussion of using Link Aggregated (LA) hybrid RF-VLC V2X communication to overcome the adverse effects of meteorological phenomena and explores its potential applications and challenges.

[1]https://lifi.co/lifi-companies/

This article is organized as follows: Section II introduces the motivation behind the potential benefit of employing hybrid VLC and RF in a vehicular communication system. Several case studies specifically relevant to link aggregated hybrid RF-VLC V2X applications under various meteorological conditions are presented in Section III. In Section IV, we outline a range of promising research directions, challenges and opportunities associated with RF-VLC in vehicular environments. Finally, concluding remarks are drawn in Section V.

## II. Hybrid RF-VLC V2X Systems

Pure RF links may suffer from excessive RF interference and limited communication resources in situations of high vehicular density. Such a situation can lead to low throughput and large communication latency owing to packet delivery failures and potentially aggressive automatic repeat request (ARQ) retransmission attempts. As a possible solution, this paper presents the core idea of leveraging the non-interfering unlicensed VLC band together with the RF band for improved V2X communications while supporting enhanced security. Furthermore, a VLC-enabled V2X system enjoys minimal setup costs as VLC-based V2X can use light-emitting diodes (LEDs) or laser diodes (LDs) that are already widely present in the vehicular head- and tail-lights or in street/traffic lights. Despite the above benefits, standalone VLC networks also have their drawbacks, including their limited transmission distance, sensitivity to background light, and line-of-sight (LOS) blockage. These impediments are conveniently circumvented by using classic RF wireless networks, which exhibit wider coverage and higher transmission reliability without LOS. By intelligently combining VLC-aided V2X communications with classic RF-based communications, it is anticipated that the overall system performance can be increased to meet the stringent requirements of 6G-V2X networks.

A typical hybrid RF-VLC based V2X systems have the potential to deliver a significantly improved vehicular message dissemination by exploiting the complementary advantages of standalone VLC and RF systems. For instance, RF can overcome the shortcomings of VLC, such as short communication range, sensitivity to ambient light and adverse weather conditions, as well as VLC's inability to propagate through opaque objects; VLC, on the other hand, can offer much higher data rates with very low interference in line-of-sight (LoS) scenarios. There are two major types of hybrid RF-VLC-based vehicular communication systems [8]:

a. **Link-Aggregated (LA) Hybrid RF-VLC V2X systems**: To improve the achievable data rate and connection reliability, the vehicular nodes communicate with both VLC and RF links simultaneously.

b. **Non-Link Aggregated (non-LA) Hybrid RF-VLC V2X systems**: In this case, the vehicular nodes utilize either VLC or RF technology at a given time to optimize the network parameters.

As illustrated in Fig. 1, there are five primary scenarios in which VLC can complement and strengthen RF commu-

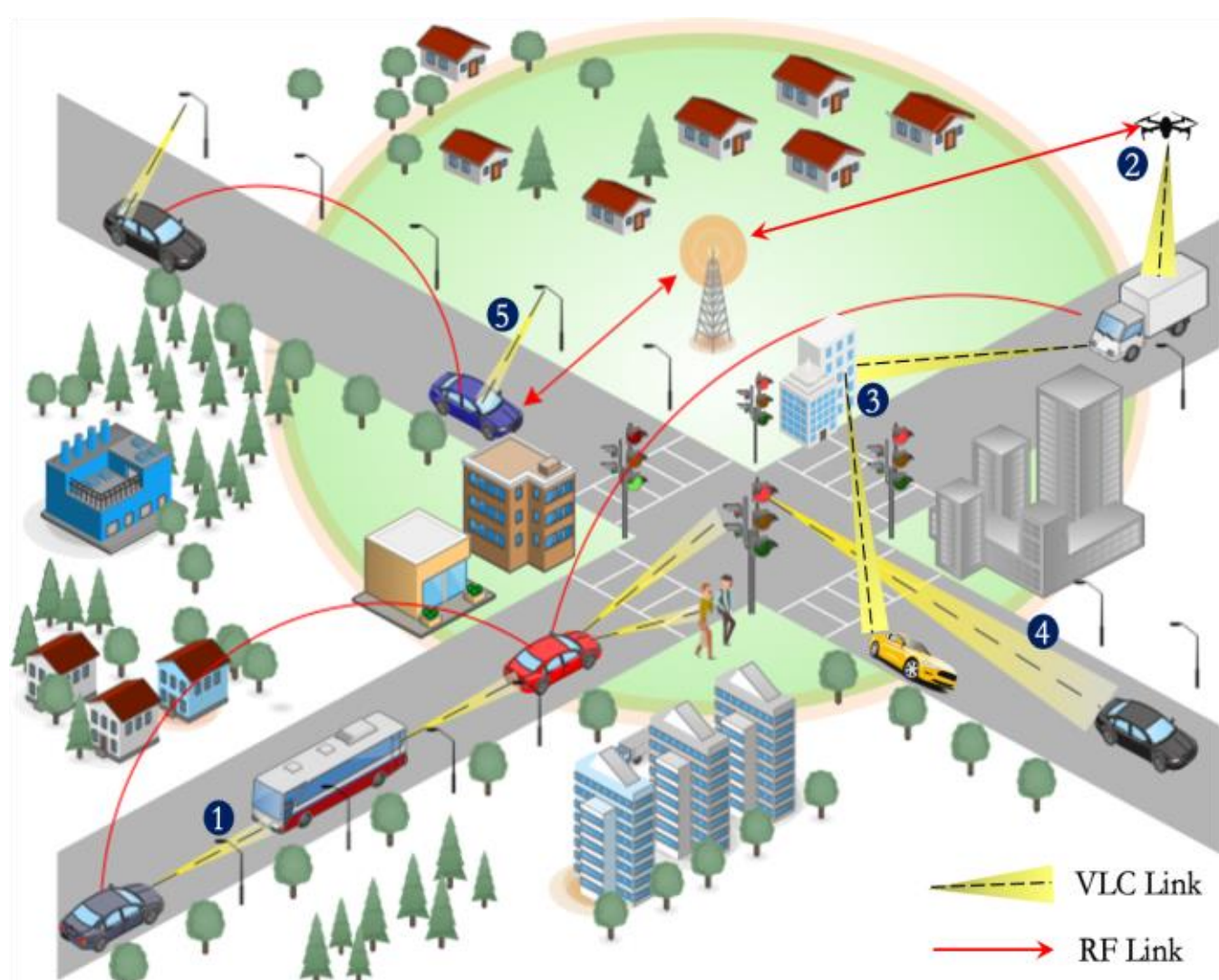

Fig. 1: Illustration of a generic hybrid RF-VLC communication in a vehicular network.

| Scenario | a. Weather Conditions | b. End-to-End Latency | c. Communication Range | d. NLOS | e. Throughput | f. Reliability |
|---|---|---|---|---|---|---|
| ① | ★★☆ ●●● | ★★★ ●●○ | ★★★ ●●● | ☆☆☆ ●●● | ★★★ ●●○ | ★★★ ●●● |
| ② | ☆☆☆ ●●● | ★★★ ●●○ | ★★☆ ●●● | ☆☆☆ ●●● | ★★★ ●●○ | ★★★ ●●○ |
| ③ | ★★☆ ●●● | ★★★ ●●○ | ★★★ ●●● | ★★★ ●●● | ★★★ ●●○ | ★★★ ●●○ |
| ④ | ★★☆ ●●● | ★★★ ●●○ | ★★☆ ●●● | ☆☆☆ ●●● | ★★★ ●●○ | ★★★ ●●○ |
| ⑤ | ★★☆ ●●● | ★★★ ●●○ | ★★☆ ●●● | ☆☆☆ ●●● | ★★★ ●●○ | ★★★ ●●○ |

☆☆☆: V-VLC does not suit. ○○○: IEEE 802.11p does not suit.
★★☆: V-VLC can be used. ●●○: IEEE 802.11p can be used.
★★★: V-VLC suits best. ●●●: IEEE 802.11p technology suits best.

TABLE I: A qualitative assessment of the capabilities of two communication technologies (i.e., V-VLC and IEEE 802.11p) in supporting five primary scenarios.

nication in V2X networks: ① **V2V communications via front or back lights**; ② **Unmanned aerial vehicle (UAV)-to-Vehicle (U2V) communication**; ③ **V2V communication via Reflecting Intelligent Surfaces (RIS)**[2]; ④ **V2X communications via traffic lights**; and ⑤ **V2X communications via street lights**. In addition to increased data rates and reduced latency, VLC has the potential to address some of the limitations of traditional RF-based V2X communication. For example, in the left bottom corner of Fig. 1, the RF-based V2V communication of two cars separated by a large bus may suffer from severe packet loss due to the shadowing effect [9]. In this scenario, the transmitting car may use VLC to communicate with the bus; subsequently, the bus could relay the messages to the receiving car in the shadowed region. Moreover, the data packets can also be relayed using traffic/street lights at urban intersections, allowing vehicles to interact across perpendicular streets, where a classical V-RF solution is often plagued by severe packet loss. Moreover, the use of optical-RIS (O-RIS)[3] can further combat packet loss, enhancing signal quality and providing wider coverage in a VLC aided V2X systems. Apart from the above, VLC-enabled UAV based U2V communication ② can be utilized for smart traffic monitoring systems to monitor, track and control the allowed speed, traffic violations and suspicious behaviours of vehicles moving on the road. Table I provides a qualitative evaluation of two communication technologies (i.e., V-VLC and IEEE 802.11p) that are capable of fulfilling (or partially) the requirements of the categorized use cases depicted in Fig. 1, based on: 1) impact factors (weather, non-LOS, distance), and 2) network quality requirements (end-to-end latency, throughput, reliability). In the next section, we briefly investigate the the potential benefits of employing the link aggregated hybrid RF-VLC V2X systems ④ against its standalone counterparts for enhanced safety message dissemination at road intersection under adverse weather conditions.

[2]RIS refers to reconfigurable metasurfaces consisting of a large number of passive antenna-elements, with which one can effectively control not only the phase, but potentially the frequency, amplitude and the polarization of the incident wireless signals [1].

[3]O-RIS can be envisioned as an extension of RIS for THz optical wireless signals and eventually for VLC [10].

## III. Case Studies

Hybrid RF-VLC is capable of significantly improving the safety at road intersections, where frequent accidents tend to occur. At road intersections, the surrounding high-rise buildings, road-side installations or sign-boards may block the LoS communication among vehicles. In this case, hybrid RF-VLC is useful for enhancing traffic safety by improving the opportunistic exchange of the V2X-specific cooperative awareness messages (CAMs) and/or basic safety messages (BSMs) among vehicles. To increase the reliability of a communication link, a relay can be placed at an intersection relying upon the existing vehicular-RF (V-RF) communication technologies. However, with the increase of vehicular density, V-RF assisted relaying may experience higher interference, lower packet reception rates, and increased communication delays due to severe channel congestion and retransmission attempts [1]. To this end, the co-deployment of vehicular-VLC (V-VLC) and V-RF communication systems is capable of improving the safety message dissemination at road intersections. Specifically, we propose to use VLC-based V2I communication ④, where the roadside unit (RSU) mounted on lamp-posts or traffic lights receive BSMs in the VLC uplink. It has been shown in [4] that non-LA hybrid RF-VLC V2X networks lead to a substantial reduction in outage along with improvements of throughput and latency as compared to pure V-VLC or pure V-RF networks.

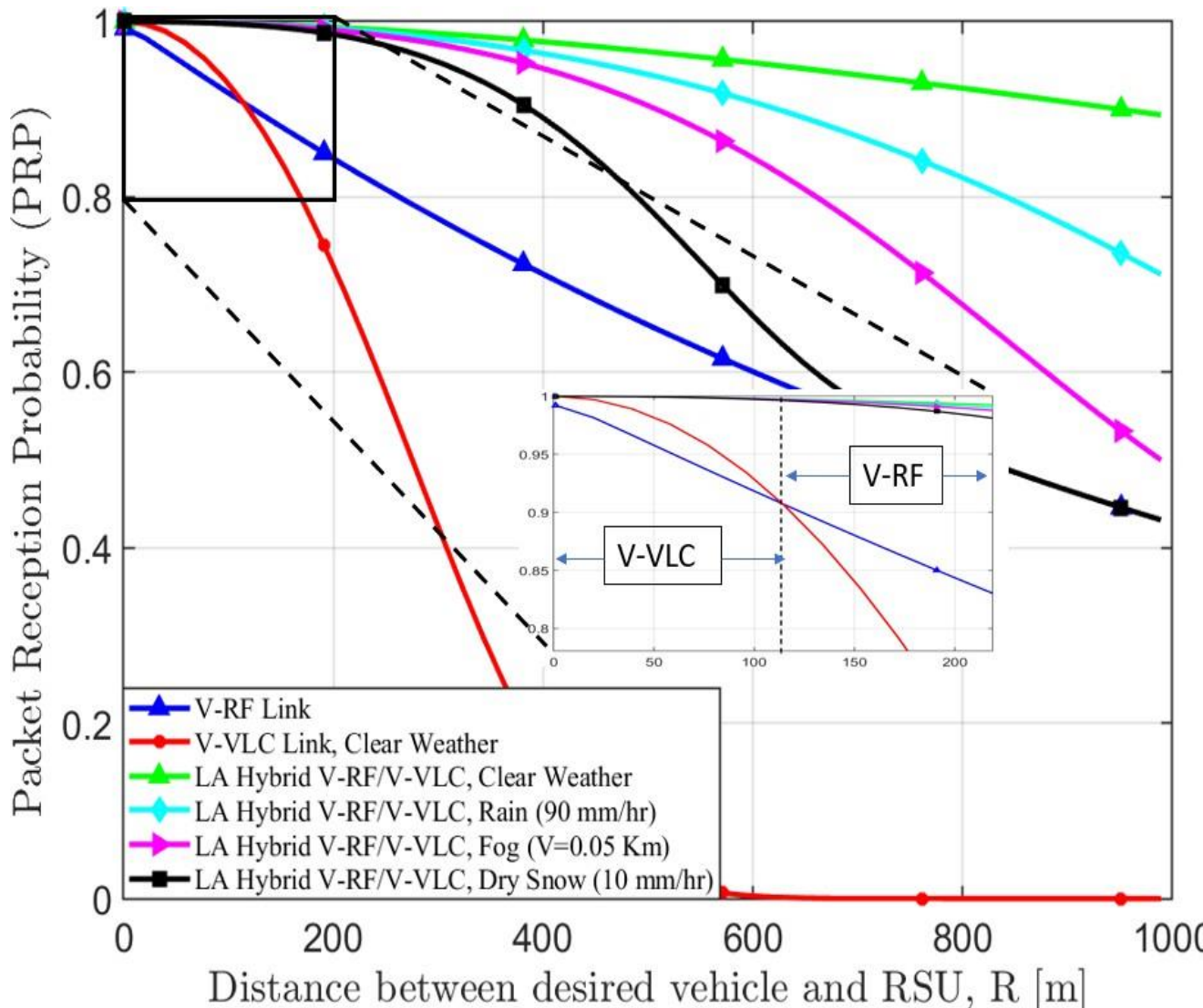

Fig. 2: PRP at RSU for pure V-RF, pure V-VLC and LA hybrid RF-VLC V2X communication systems under rain, fog and dry snow conditions.

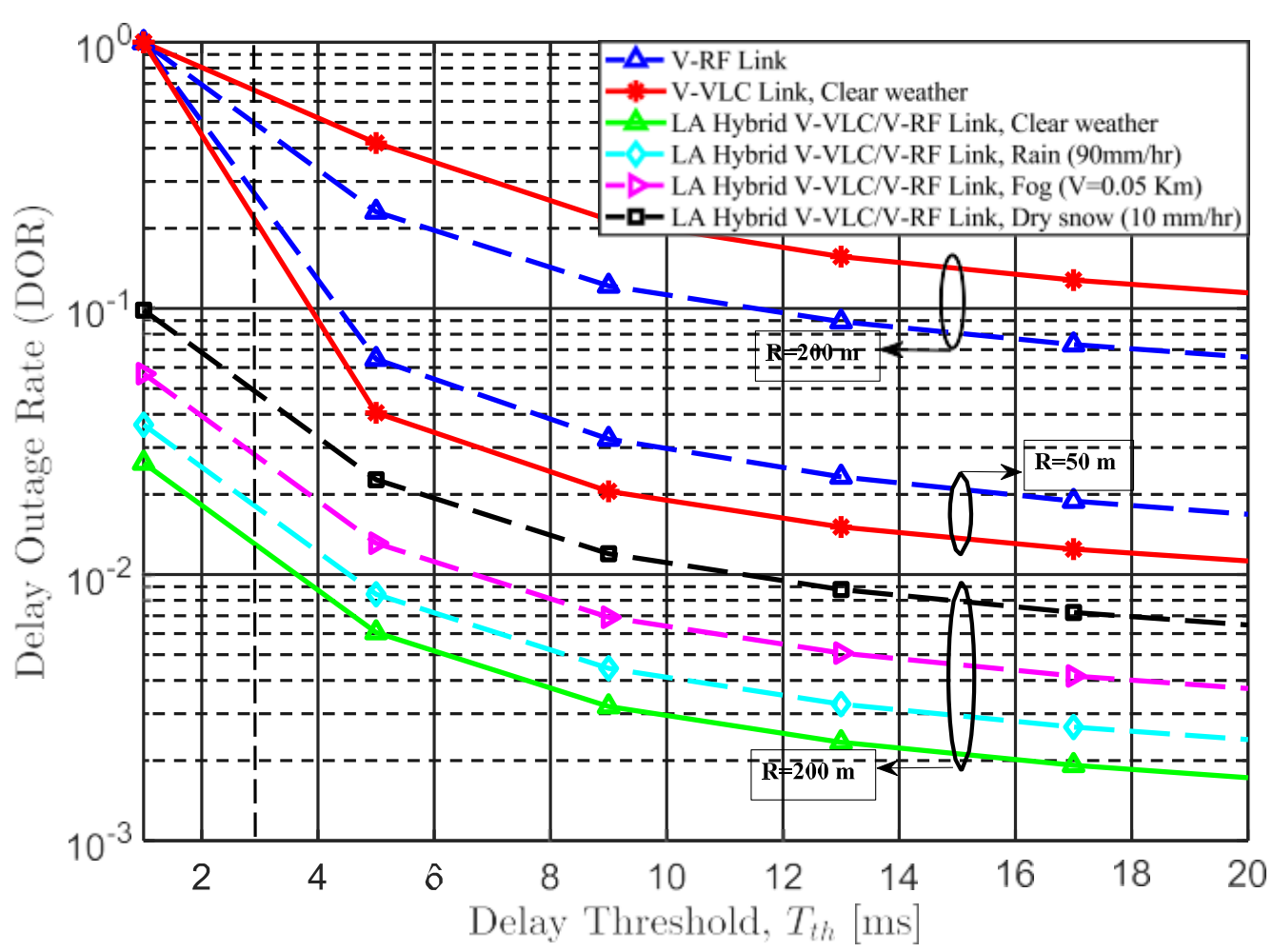

Fig. 3: Delay outage performance for pure V-RF, pure V-VLC and LA hybrid RF-VLC V2X communication system as a function of delay threshold, $\square_{\square h}$.

As an illustration of a quantitative analysis, we consider a typical road intersection scenario ④. We assume that the desired and interfering vehicles are equipped with both VLC and RF transceivers. The desired vehicle close to intersection is assumed to carry critical road information (for instance, such as the vehicle position, speed, acceleration and flow of direction) which needs to be immediately communicated to the RSU assumed to be positioned at the center of the road intersection. Now, we consider a LA hybrid RF-VLC V2X uplink scenario and compare its performance to that of a pure V-VLC and pure V-RF uplink, taking into consideration the impact of interference and various meteorological phenomena such as rain, fog, and dry snow conditions. We use a stochastic geometry-based modelling approach for the analysis. The LA technique enhances the total available bandwidth and leads to more reliable network performance as well as a reduction in the end-to-end latency as demonstrated in [11]. We consider that the communications between the RSU and the desired vehicle are subject to interference from vehicles in the same lane and also from vehicles in perpendicular lanes. The system parameters were chosen in accordance with a practical vehicular communication scenario. We performed Monte Carlo (MC) simulations leveraging stochastic geometry-based modelling as described in [4]. Note that MC simulations are obtained by averaging over 10,000 realizations of the Poisson point process (PPP) and fading parameters to obtain reliable statistics of the interference. It may be noted that the vehicles are deployed over a length of 10,000 m, and the interference originating from interfering vehicles transmitting concurrently is summed and stored. In addition, the attenuation coefficient under rain (rain rate=90 mm/hr), fog (V=0.05 Km) and dry snow (snow rate=10 mm/hr) are taken to be 21.9, 78.8, and 131 dB/km, respectively, as given in [7, Table 2]. Unless otherwise stated, we assume having the vehicular density □ and channel access probability □ to be 0.01 and 0.01, respectively. Observe from Fig. 2 that depending on the transmitter's location, the pure V-VLC and V-RF systems exhibit complementary roles in terms of packet reception probability (PRP). For MC based system level modelling, we have adopted Lambertian channel model for vehicular-VLC system. The VLC channel gain is strongly dependent on the angle of irradiance of LED transmitter, the angle of incidence/ field-of-view (FoV) of optical receiver and the distance between transmitter and receiver. For the given angle of irradiance and receiver's FoV, the received optical power is more since VLC link's coherence time is relatively larger than RF coherence time for smaller distance between desired vehicle and RSU. Additionally, the impact of interference from adjacent lane vehicles is also less pronounced due to restricted receiver's FoV. As a consequence, the packet reception probability associated with standalone VLC is more compared to standalone RF for the proposed framework, specifically when the typical distance between the desired vehicle and the RSU is less than 120 m. In contrast, with increase in distance between desired vehicle and RSU, the received optical power reduces drastically (due to cumulative effect of large distance and interference). In such a case, V-RF communication is a more feasible communication means between the desired vehicle and RSU. Interestingly, regardless of the distance between the desired vehicle and the RSU and any prevailing weather conditions, the LA hybrid RF-VLC V2X system outperforms the pure V-VLC or V-RF links. Notice that the performance of the LA system is strongly negatively influenced by dry snow conditions. This is primarily due to high attenuation in the pure V-VLC link under dry snow conditions.

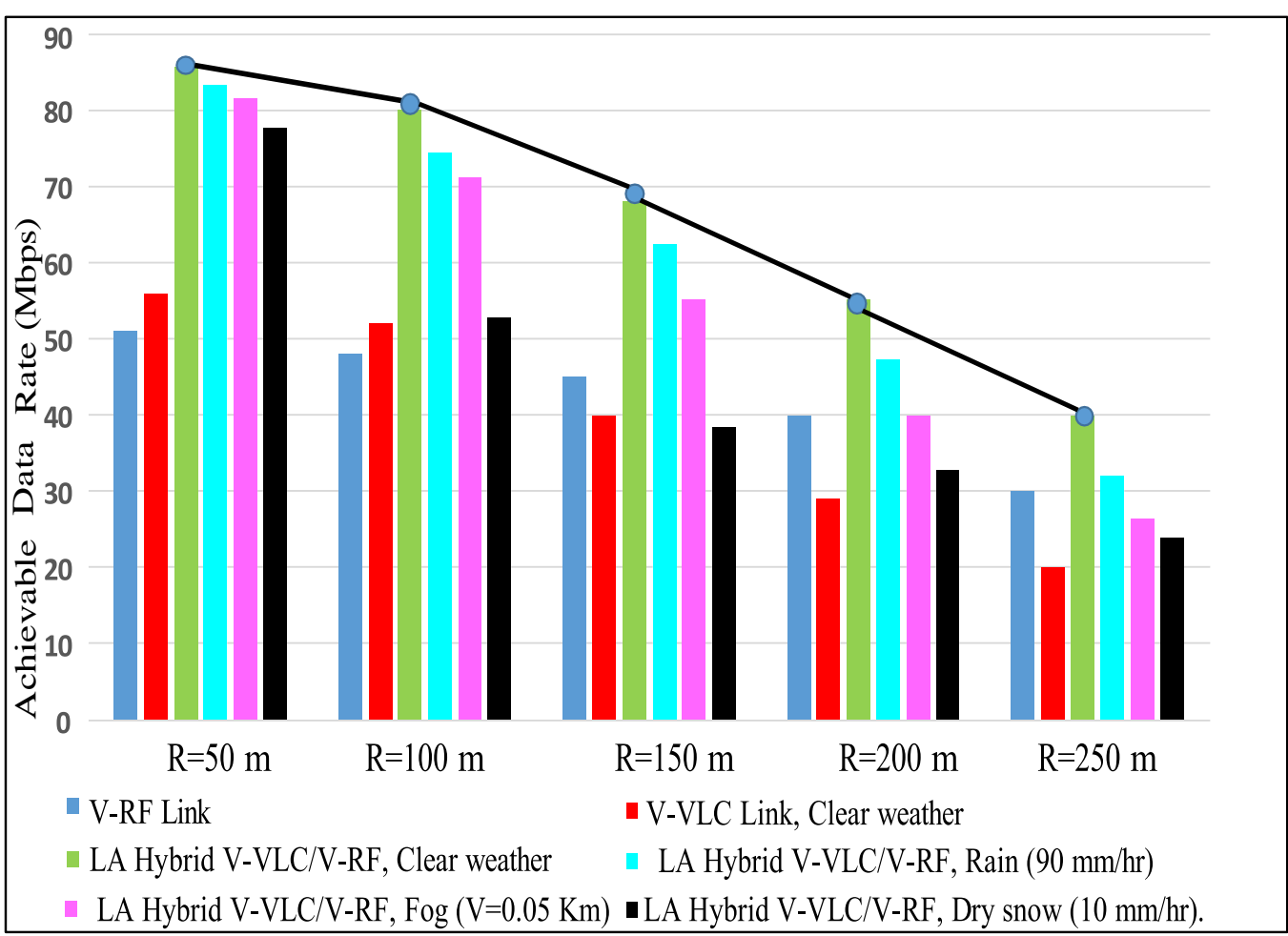

Fig. 4: Achievable data rate for different network configurations for different values of distance between RSU and the desired vehicle.

Many warning/safety-specific messages are life-critical, hence, high latency is unacceptable, especially in accident-prone situations. It is anticipated that the hybrid RF-VLC V2X systems can offer ultra-reliable low latency communication (URLLC) among vehicles while meeting 6G key performance indicators (KPIs) for vehicular network requirements. According to [4, Eq.(28)], we consider the metric of delay outage rate (DOR), which represents the probability that the minimum transmission time (MTT) required for sending a certain amount of data is higher than the tolerable duration. We plot the DOR of standalone V-RF, LA hybrid RF-VLC V2X, and pure V-VLC guaranteeing different maximum delay requirements for both 50□ and 200□ distances in Fig. 3. Here, we assume that the system bandwidth for the pure RF and VLC systems is 20 MHz[4]. Again, depending on the transmitter's location, pure V-VLC and V-RF exhibit complementary roles, as evidenced by Fig. 3. Additionally, for data traffic having stringent delay requirements of $< 1$ ms, the LA-aided hybrid RF-VLC V2X under any weather condition ensures having the minimum delay in transmitting data size, □=50KB from the desired vehicle to the RSU as compared to pure V-VLC or V-RF systems. In light of the above results, it can be inferred that irrespective of the meteorological phenomena, the LA-aided hybrid RF-VLC V2X system achieves ultra high reliability (∼99.999%) and ultra-low latency (<1 ms) up to □=200m. For an interference-limited scenario, the LA-aided hybrid RF-VLC V2X system meets stringent reliability and latency requirements for advanced vehicular scenarios [2]. The achievable rate associated with link aggregated hybrid RF-VLC employing a LA-SINR algorithm is governed by [11, Eq.(13)]. Fig. 4 shows the achievable data rate for different network configurations of □∈{50□, 100□, 150□, 200□, 250□}. Here, we assume that the desired vehicle accesses the channel at a transmission probability of □$_□$=0.9 and the link aggregation overhead is □$_{□□}$=0.8. In an interference-limited scenario, the maximum achievable data rate associated with LA-aided hybrid RF-VLC V2X can be as high as 83.2 Mbps at □=50m, which reduces to 39.8 Mbps at □=250m under clear weather conditions. Note that the achievable rate of a non-LA hybrid RF-VLC V2X system relies on the maximum throughput offered by either pure V-VLC or pure V-RF systems. The results presented meet the data rate requirements of advanced vehicular scenarios that have been suggested in [2]. Note that the advanced driving scenario includes semi-automated or fully-automated driving use cases for longer inter-vehicle distances, supporting data rates ranging from 10-53 Mbps among vehicles or RSUs in close proximity.

[4]RF spectrum is generally licensed and expensive, whereas VLC spectrum is free, hence more cost-effective.

## IV. Challenges and Future Directions

In this section, we present several potential applications and research areas for hybrid RF-VLC V2X communication systems. Fig. 5 illustrates an architecture for a hybrid RF-VLC V2X network, which is based on three layers: infrastructure, network, and application. The interoperability among different layers caters to realize the following applications and also reveal some promising research directions associated with hybrid RF-VLC V2X networks.

### *A. Reconfigurable Intelligent Surfaces (RIS) Enabled Mixed RF-VLC V2X Systems*

Recently, RIS has attracted much research attention owing to its salient feature of transforming hostile wireless channels into benign ones. It has emerged as a disruptive communication technology for enhancing signal quality and transmission coverage in wireless vehicular networks. Significant gains may be gleaned by incorporating RISs into hybrid RF-VLC V2X systems (3) thanks to the enhanced resilience to LoS blockages, especially when striking a flexible tradeoff between light and RF communications, quality-of-service, interference mitigation, enhanced localization services, and improved energy harvesting.

In particular, 6G-V2X can benefit from RISs in situations where coverage is constrained. For instance, in urban areas, road intersections **(d, Table I)** constitute an ideal use case for deploying RIS-aided RF-VLC V2X systems, where the exchange of safety messages between vehicle lights may be blocked by buildings, walls, surrounding vehicles and other obstructions, as shown in Fig. 5**A**. Indeed, RIS-aided hybrid RF-VLC V2X systems offers a promising societal sustainable solution for road intersection scenarios by efficient safety message information exchange, thereby avoiding traffic accidents and improving road safety. By enabling an RIS controller to actively relay the information from the source to the destination vehicle, the RIS can not only potentially help improve the transmission rate for standalone V-VLC systems, but also provide a wide coverage range using RIS controller. As soon as the quality of VLC link degrades (eg. long-range communication case), the communication between source and destination vehicles can still be accomplished using conventional V-RF systems employing relaying. Nonetheless, several distinctive research challenges such as channel estimation in highly dynamic scenarios, optimal RIS deployment, reliable energy management schemes, optimal resource allocation and reflection optimization have to be carefully addressed before the practical integration of RIS into hybrid RF and VLC vehicular communication systems.

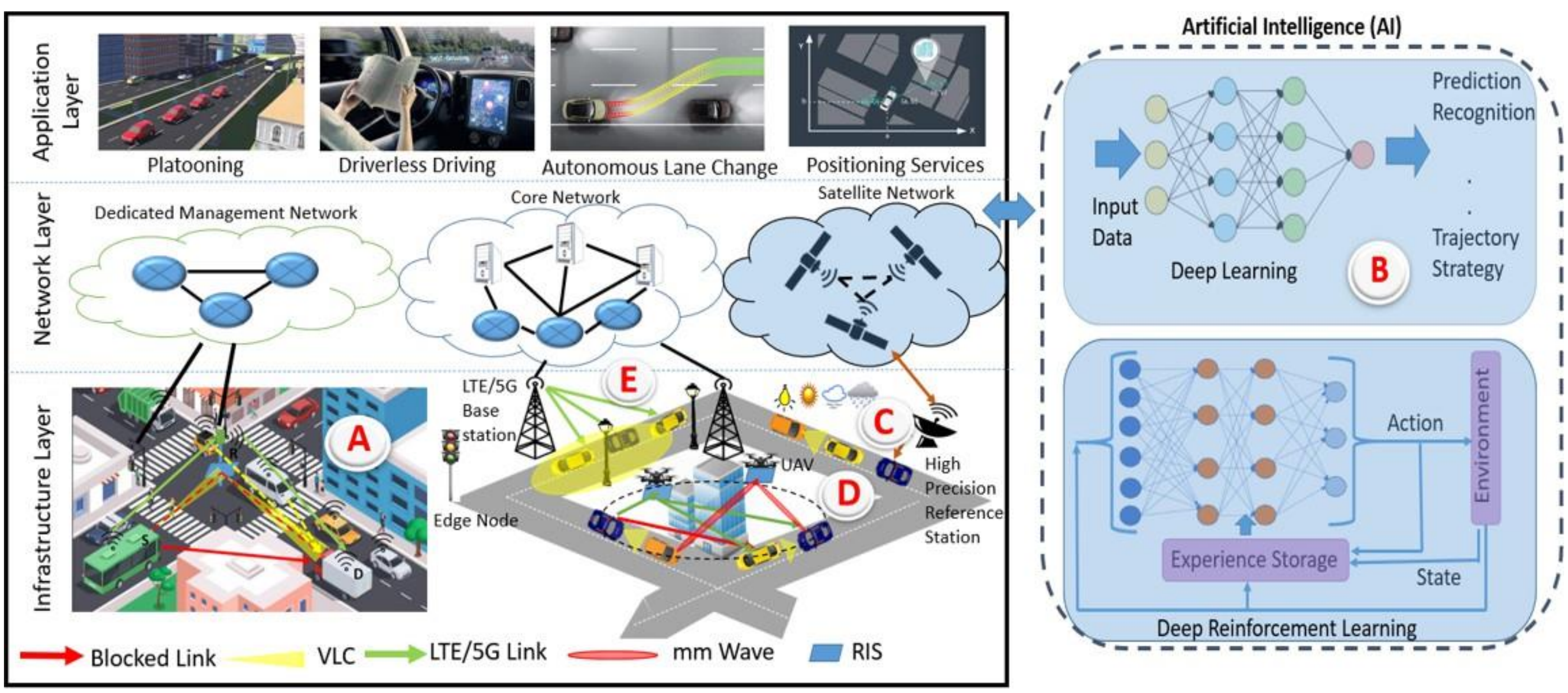

Fig. 5: Envisioned use-case scenarios and possible research areas for hybrid RF-VLC V2X communication systems.

### B. AI-empowered System Design

Artificial Intelligence (AI) assisted 6G is expected to unlock the promise of future V2X [12]. These features are desirable in vehicular networks to accommodate diverse and advanced use cases and their technical requirements. In fact, vehicular networks with AI based self-regulation can detect anomalies in the network and lower output by modifying the power. AI-driven energy management will reduce costs and energy consumption in networks. AI-powered vehicle networks may lead to fully autonomous V2X operations in the future, thereby reducing energy consumption and carbon emissions. Due to vehicular networks' inherent heterogeneity and mobility, communication environments are highly complex, resulting in varying wireless or optical channels. On the other hand, different layers of the current RF and VLC communication systems are optimized independently. Such a design paradigm may not be ideal when dealing with diverse quality of service (QoS) requirements **(b, e, f, Table I)**, particularly when dealing with complex and dynamic vehicular environments. There is a need to configure different functional blocks of VLC/RF communication systems in a joint and adaptive manner according to the dynamically varying vehicular network. For example, Machine learning (ML)-assisted adaptive coding and modulation (ACM) is expected to improve the robustness whilst reducing the communication latency. ML can also be applied to optimize multiple configurations simultaneously. An end-to-end hybrid communication architecture needs to be considered when an ML-based joint optimization is developed. A hybrid RF-VLC V2X system would also face resource allocation issues such as bandwidth allocations and access point selection based on the requirements of the network, availability of resources, and mobility of the vehicles. In addition, dynamic decision-making on whether to use LA or non-LA hybrid techniques can be crucial for effective and energy-efficient V2X communications. Using traditional methods of resource allocation would mean re-running the simulation for every small change, resulting in significantly large overhead [12]. In such a case, ML-based techniques can be an effective tool for data-driven decision-making in order to improve the performance of resource allocation in RF/VLC vehicular networks. In particular, a reinforcement learning solution for hybrid RF-VLC V2X systems can be helpful in tackling the challenges arising due to dynamic environments and shortage of relevant datasets for vehicular networks. Future research may be devoted to developing ML-based resource allocation algorithms for RF/VLC V2X network with the goal of ensuring maximum network performance and decrease in control overhead and handover latency.

### C. Deployment Issues

When it comes to a sustainable transportation, hybrid RF-VLC V2X systems offers an effective solution as it contributes directly to a reduction in carbon footprints and thus has a positive impact on environmental sustainability. Despite the immense potential of hybrid RF-VLC V2X systems, their widespread deployment can be hampered by the availability of VLC links under meteorological phenomena such as rain, fog, snow and hazy conditions **(a, Table I)** [7]. In addition, solar irradiance, road conditions, and artificial light sources (e.g., roadside illumination, sign boards, fluorescent lamps) also pose challenges for such hybrid systems in the real world. Further, the intensity of the signal received in VLC can vary widely due to the vehicle's mobility. Thus, channel variations caused by mobility and ambient light-induced noise must be carefully addressed before implementing VLC in the 6G-V2X ecosystem. Compared with V-RF, V-VLC are subject to light-path blockages, which can drastically reduce the data rate in such hybrid vehicular applications. The authors of [13] overcome this challenge by proposing omnidirectional and ubiquitous coverage in VLC. It would also be interesting to examine how the hybrid system can cope with dynamic changes in the vehicular environment, such as mobility, topology, traffic density, and user demand. Furthermore, the specific bandwidth aggregation in LA-aided hybrid RF-VLC V2X systems constitutes an open research challenge, given for example 1Hz RF bandwidth in the sub-6GHz band and 1Hz VLC bandwidth in the 800 THz band. In light of the above discussions, it is clear that these challenges have to be tackled before the practical deployment of such hybrid systems.

### D. Coexistence of mmWave, THz and VLC

Both VLC and TeraHertz (THz) techniques constitute promising candidates for realizing the vision of 6G V2X. It is anticipated that the operation of 6G V2X will rely on usage of a wide range of transmission bands including RF, VLC, THz, and mm-wave frequencies. There exists a trade-off among coverage area, ergodic rate, mobility and latency when dealing with a broad range of radio/optical spectrum **(b, c, d, e, f, Table I)**. There can be two ways to realize the presence of multiple frequencies: flexible spectrum coexistence and hybrid deployment. In the flexible spectrum coexistence approach, the base stations (BSs) with different frequencies are deployed separately and each BS at a certain time can operate on only one of RF, VLC, and THz frequency bands. For flexible multi-band utilization, the multi-band C-V2X system needs advanced front-end hardware. In addition, the coexistence of different network spectra leads to new interference problems [14]. In the hybrid approach, each BS relies on more than one frequency band. Optimizing user-side opportunity spectrum selection, network activation mechanisms considering traffic load, BS implementation, and multichannel solutions will be the key challenges for such multiband vehicular networks (MBVNs).

### E. NOMA and its variants

Multiple access plays a pivotal role in vehicular communication and networking. In DSRC, carrier sense multiple access (CSMA) is adopted, whereby all vehicles that wish to communicate must constantly sense the availability of the channel. CSMA is simple, yet may lead to large communication overhead and high collision rates in dense vehicular networks. LTE-V2X and 5G-NR-V2X, on the other hand, use OFDMA for multiple access, but may suffer from the same problem in modes operating without BS assistance. Given the explosive growth of communication sensors and connected vehicles, much research has been carried out in recent years on non-orthogonal multiple access (NOMA) for supporting massive numbers of concurrent communication links. Both power-domain NOMA and code-domain NOMA (e.g., sparse code multiple access) may be applied to hybrid RF-VLC based V2X communication systems. K. G. Rallis *et al.* [15] investigated a cross-band aggregated VLC/RF NOMA network with an RF relay link from a cell-center user to a cell-edge one, aiming to improve the latter's performance with the assumption that the cell-edge user receives its message from both VLC and RF links in an aggregated manner. Simulation results show the superiority of the cooperative NOMA protocol versus benchmarks that do not use NOMA and/or cooperative sidelink communications. In this line of research, it is interesting to optimize the user grouping, power allocation, codebook design, and multiuser detection algorithms to meet the diverse QoS requirements of future vehicular networks. Furthermore, a cross-layer design of NOMA systems involving both the physical layer and the medium access layer is still missing. As NOMA can support a higher number of concurrent communication links, an enhanced re-transmission protocol (unlike the one for traditional orthogonal multiple access systems) that considers the unique characteristics and channel conditions of V-VLC and V-RF is desirable.

## V. Concluding remarks

This article provides useful insights on leveraging hybrid RF-VLC technology for enhanced vehicular applications. The MC simulation results demonstrate that our proposed LA-aided hybrid RF-VLC V2X systems have the capabilities to provide a highly reliable and low-latency V2X communication, even in scenarios affected by interference and adverse meteorological conditions. Further, the maximum achievable data rate associated with the proposed system can be as high as 78 Mbps at 50 m. Such a rate reduces to 34 Mbps at 200 m under dry snow conditions, which is primarily due to high attenuation in the pure V-VLC link under dry snow conditions. Finally, we discussed the potential role of RIS and ML in such hybrid RF-VLC V2X systems along with several key research challenges and interesting open issues. In this article, we have identified the challenges associated with practical deployment of such hybrid systems such as channel variations due to mobility, efficient bandwidth aggregation, adverse weather conditions and ambient light induced interference. We have pointed out that VLC and THz are spectra with highly promising prospects to realize the vision of 6G-V2X. We also believe that NOMA is a very promising candidate technique for the next generation 6G-V2X networks mainly due to massive concurrent connections and increased spectrum efficiency. It is hoped that this article will provide useful guidance for future research in this emerging and promising area of hybrid RF-VLC V2X systems, paving the way to a more economic, societal and environment sustainable future solution for 6G based V2X era.

## References

[1] M. Noor-A-Rahim, Z. Liu, H. Lee, M. O. Khyam, J. He, D. Pesch, K. Moessner, W. Saad, and H. V. Poor, "6G for vehicle-to-everything (V2X) communications: Enabling technologies, challenges, and opportunities," *Proceedings of the IEEE*, vol. 110, no. 6, pp. 712–734, 2022.

[2] D. P. Moya Osorio, I. Ahmad, J. D. V. Sánchez, A. Gurtov, J. Scholliers, M. Kutila, and P. Porambage, "Towards 6G-enabled internet of vehicles: Security and privacy," *IEEE Open J. Commun. Soc.*, vol. 3, pp. 82–105, 2022.

[3] T. T. Sarı, M. K. Assoy, and G. Seçinti, "Utilizing smartphones for blind spot detection," in *Proc. IEEE 20th Consumer Communications Networking Conference (CCNC)*, 2023, pp. 911–912.

[4] G. Singh, A. Srivastava, V. A. Bohara, Z. Liu, M. Noor-A-Rahim, and G. Ghatak, “Heterogeneous visible light and radio communication for improving safety message dissemination at road intersection,” *IEEE Trans. Intell. Transport. Syst.*, pp. 1–13, 2022.

[5] A. Memedi and F. Dressler, “Vehicular visible light communications: A survey,” *IEEE Commun. Surveys Tut.*, vol. 23, no. 1, pp. 161–181, 2021.

[6] C. Lee, C. Tan, H. Wong, and M. Yahya, “Performance evaluation of hybrid VLC using device cost and power over data throughput criteria,” in *Ultrafast Imaging and Spectroscopy*, vol. 8845. Proc. SPIE, 2013, pp. 130–139.

[7] G. Singh, A. Srivastava, and V. A. Bohara, “Impact of weather conditions and interference on the performance of VLC based V2V communication,” in *Proc. IEEE Intl. Conf. Transparent Optical Net. (ICTON)*, 2019, pp. 1–4.

[8] H. Abuella, M. Elamassie, M. Uysal, Z. Xu, E. Serpedin, K. A. Qaraqe, and S. Ekin, “Hybrid RF/VLC systems: A comprehensive survey on network topologies, performance analyses, applications, and future directions,” *IEEE Access*, vol. 9, pp. 160 402–160 436, 2021.

[9] H. Nguyen, X. Xiaoli, M. Noor-A-Rahim, Y. L. Guan, D. Pesch, H. Li, and A. Filippi, “Impact of big vehicle shadowing on vehicle-to-vehicle communications,” *IEEE Trans. Veh. Tech.*, vol. 69, no. 7, pp. 6902–6915, 2020.

[10] S. Aboagye, A. R. Ndjiongue, T. M. N. Ngatched, O. A. Dobre, and H. V. Poor, “RIS-assisted visible light communication systems: A tutorial,” *IEEE Commun. Surveys Tut.*, vol. 25, no. 1, pp. 251–288, 2023.

[11] N. M. Karoti, S. Paramita, R. Ahmad, V. A. Bohara, and A. Srivastava, “Improving the performance of heterogeneous LiFi-WiFi network using a novel link aggregation framework,” in *Proc. IEEE Wireless Communications and Networking Conference (WCNC)*, 2022, pp. 2322–2327.

[12] L. Liang, H. Ye, and G. Y. Li, “Toward intelligent vehicular networks: A machine learning framework,” *IEEE Internet of Things Journal*, vol. 6, no. 1, pp. 124–135, 2019.

[13] H. B. Eldeeb, S. M. Sait, and M. Uysal, “Visible light communication for connected vehicles: How to achieve the omnidirectional coverage?” *IEEE Access*, vol. 9, pp. 103 885–103 905, 2021.

[14] T. Xu, M. Zhang, Y. Zeng, and H. Hu, “Harmonious coexistence of heterogeneous wireless networks in unlicensed bands: Solutions from the statistical signal transmission technique,” *IEEE Veh. Tech. Mag.*, vol. 14, no. 2, pp. 61–69, 2019.

[15] K. G. Rallis, V. K. Papanikolaou, P. D. Diamantoulakis, S. A. Tegos, A. A. Dowhuszko, M.-A. Khalighi, and G. K. Karagiannidis, “Energy efficient cooperative communications in aggregated VLC/RF networks with NOMA,” *IEEE Trans. Commun.*, vol. 71, no. 9, pp. 5408–5419, 2023.

**Gurinder Singh** (gurinder.singh@lnmiit.ac.in) is currently an Assistant Professor with Department of Communication and Computer Engineering, the LNM Institute of Information Technology (LNMIIT), Jaipur, India. His research interests include vehicular-visible light communication, hybrid VLC-RF architecture, non-orthogonal multiple access (NOMA), and reconfigurable intelligent surfaces (RIS)-aided vehicular communication systems.

**Anand Srivastava** (anand@iiitd.ac.in) is the Vice Chancellor at Netaji Subhas University of Technology (NSUT), New Delhi, India. He is M.Tech. and Ph.D. from Indian Institute of Technology (IIT) Delhi. Prior to his appointment at NSUT, he adorned the position of Professor in the Department of Electronics and Communication Engineering (ECE) at the illustrious Indraprastha Institute of Information Technology (IIIT), Delhi, India and Director at IIIT Delhi Incubation Center (a Section 8 company). Earlier, he had 20 years of experience with the Center for Development of Telematics (CDOT), a telecom research center of Govt. of India where he was involved in the development of national-level telecom projects. His research work is in the area of optical networks, vehicle-to-vehicle communications, Fiber-Wireless architectures, and Visible Light Communications.

**Vivek Ashok Bohara** (vivek.b@iiitd.ac.in) is a Professor with Department of Electronics and Communication Engineering, Indraprastha Institute of Information Technology (IIIT), Delhi, India. He obtained his PhD while working with Centre for Infocomm Technology within the school of Electrical and electronic engineering in Nanyang technological university (NTU), Singapore. His research interests are next generation communication technologies such as Visible Light Communication (VLC), hybrid RF-VLC communication, integration of optical communication with intelligent reflective surfaces (IRS), UAV and vehicular communication.

**Md. Noor-A-Rahim** (m.rahim@cs.ucc.ie) is currently a Lecturer (Assistant Professor) with the School of Computer Science and IT, University College Cork, Cork, Ireland. His research interests include Wireless Networks, Intelligent Transportation Systems, Machine Learning, and Signal Processing. For more details, please visit: https://narahim.github.io/

**Zilong Liu** (zilong.liu@essex.ac.uk) is currently a Senior Lecturer (Associate Professor) with the School of Computer Science and Electronics Engineering, University of Essex. His research lies in the interplay of coding, signal processing, and communications, with a major objective of bridging theory and practice as much as possible. He is an Associate Editor of IEEE Transactions on Neural Networks and Learning Systems, IEEE Transactions on Vehicular Technology, IEEE Wireless Communications Letters, and IEEE Access. More details of his research can be found in the link here: https://sites.google.com/site/zilongliu2357

**Dirk Pesch** (dirk.pesch@ucc.ie) is a Professor with the School of Computer Science and Information Technology, University College Cork, Ireland. He was previously the Head of the Nimbus Research Centre, at Munster Technological University. His research interests include problems associated with architecture, design, algorithms, and performance evaluation of low power, dense, and vehicular wireless/mobile networks and services for Internet of Things and cyber-physical system’s applications in building management, smart connected communities, independent living, and smart manufacturing.